\documentclass[11pt]{article}

\usepackage[final]{latex/acl}

\usepackage{times}
\usepackage{latexsym}

\usepackage[T1]{fontenc}
\usepackage[utf8]{inputenc}

\usepackage{microtype}

\usepackage{inconsolata}

\usepackage{graphicx}

\usepackage{tcolorbox}
\usepackage{hyperref}
\usepackage{booktabs}
\usepackage{multirow}
\usepackage{verbatim}
\usepackage{fancyvrb}
\usepackage{fvextra}
\usepackage{amsmath}

\title{The Effects of Structured LLM-Generated Feedback on Programming Assignment Performance}

\author{
 \textbf{Tsvetomila Mihaylova\textsuperscript{1}},
 \textbf{Evanfiya Logacheva\textsuperscript{1}},
 \textbf{Arto Hellas\textsuperscript{1}},
 \textbf{Jing Fan\textsuperscript{1}},
\\
 \textbf{Francisco Castro\textsuperscript{2}},
 \textbf{Bita Akram\textsuperscript{3}},
 \textbf{Narges Norouzi\textsuperscript{4}},
 \textbf{Peter Brusilovsky \textsuperscript{5}},
 \textbf{Juho Leinonen\textsuperscript{1}}
\\
 \textsuperscript{1}Aalto University, Department of Computer Science,
 \textsuperscript{2}New York University,
\\
 \textsuperscript{3}North Carolina State University,
 \textsuperscript{4}UC Berkeley,
 \textsuperscript{5}University of Pittsburgh
\\
 \small{
   \textbf{Correspondence:} \href{mailto:tsvetomila.mihaylova@aalto.fi}{tsvetomila.mihaylova@aalto.fi}
 }
}

\begin{document}

% \title{The Effects of Structured LLM-Generated Feedback on Programming Assignment Performance}
% \titlenote{(Does NOT produce the permission block, copyright information nor page numbering). For use with edm\_article.cls.}}

% Submissions for EDM are double-blind: please do not include any author names or affiliations in the submission. 
% Anonymous authors:
% \numberofauthors{9}
% \author{
% Anonymous\\
%        \affaddr{Anonymous Institution}\\
%        \email{anonymous@anonymous.edu}
% }

\maketitle
% Commands
\newcommand{\dbbeforefeedback}{\texttt{Before\_Feedback}}
\newcommand{\dbafterfeedback}{\texttt{After\_Feedback}}

\newcommand{\hintadvanced}{\texttt{Diagnostic Feedback}}
\newcommand{\hintnovice}{\texttt{Single-Issue Feedback}}
\newcommand{\hintgeneral}{\texttt{General Feedback}}
\newcommand{\hintnohint}{\texttt{No Feedback}}

\newcommand{\novicestudents}{\texttt{Novice Students}}
\newcommand{\advancedstudents}{\texttt{Advanced Students}}

\newcommand{\votegood}{\texttt{Good}}
\newcommand{\votebad}{\texttt{Poor}}

\newcommand{\coursepartoldcontent}{\texttt{Old Content}}
\newcommand{\coursepartnewcontent}{\texttt{New Content}}

\newcommand{\assignmentcompletion}{\texttt{Assignment Completion}}
\newcommand{\timetosuccess}{\texttt{Time to Success}}
\newcommand{\attemptstosuccess}{\texttt{Attempts to Success}}

\newcommand{\ie}{\textit{i.e.}}
\newcommand{\ex}{\textit{ex.}}

\newtcolorbox{promptbox}[2][]{
  enhanced,
  sharp corners,
  boxrule=0.5pt,
  colback=white,
  colframe=black!60,
  title={#2},
  fonttitle=\bfseries\footnotesize,
  left=1mm,right=1mm,top=0.8mm,bottom=0.8mm,
  #1
}

\begin{abstract}
When programming students encounter errors in their code, compiler messages or static analysis output often provide limited guidance, particularly for novice programmers. Personalized feedback from instructors can be effective but does not scale well. Recent advances in large language models (LLMs) enable automated feedback generation at scale.
This study examines whether LLM-generated feedback with different levels of guidance is associated with differences in students’ problem-solving behavior. We analyze effects on time to solution and number of attempts, and examine whether these effects differ by programming experience.
We design three feedback types and compare them to a baseline in which students receive only compiler error messages.
Results from an online programming course show that LLM-generated feedback is associated with faster time to solution compared to the no-feedback baseline, with less guided feedback showing slightly stronger effects. 
Overall, the findings suggest that feedback structure plays an important role in how students progress toward correct solutions and motivate further work on adaptive feedback designs and longer-term learning outcomes.
\end{abstract}

\section{Introduction}

When students submit incorrect programming solutions, they typically receive feedback in the form of compiler errors or failed test cases \cite{10.1145/3231711}. Such feedback indicates that a solution is incorrect but often provides limited guidance on how to proceed, particularly for less experienced programmers \cite{becker2016effective}. Human instructors can offer more targeted feedback, but this does not scale to large courses \cite{messer2024grading}.
Recent work explored the use of large language models (LLMs) to generate programming feedback automatically \cite{hellas2023exploring,10.1145/3723010.3723034,vassar2024finetuning,10.1145/3657604.3662040,koutcheme2024open,koutcheme2025evaluating}. Prior studies report that LLMs can produce readable explanations and hints, but also note that generic LLM-generated feedback may be overly detailed, provide complete solutions, or contain inaccuracies \cite{xiao2024exploring,leinonen2023using,brown2025howzat}. Research in computing education further shows that novice and more experienced programmers differ in how they approach programming problems and respond to guidance \cite{chuang2024analyzing,kalyuga2007expertise}.

This study examines how the structure of LLM-generated programming feedback is associated with student performance in an online programming course. Rather than focusing on fully personalized or adaptive feedback, we compare feedback formats that differ in the amount of guidance and information they provide, ranging from more detailed explanations to more constrained, diagnostic hints. We evaluate how these different feedback structures relate to students’ time to solution and number of attempts, and examine whether their effects vary across students with different programming experience. In addition, we investigate whether the effectiveness of LLM-generated feedback differs across course content that varies in recency relative to the model training data, which may affect feedback quality.

We address the following research questions:
% \begin{enumerate}
    \textbf{RQ1}: How does the different feedback complexity affect students’ performance, in terms of (a) the expected number of attempts and (b) the time required to solve an assignment?
    \textbf{RQ2}: 
    Do the effects of different hint types on attempts and solution time differ between students with different levels of expertise?
    \textbf{RQ3}:
    How does the recency of the instructional materials relate to students’ performance, as measured by the expected number of attempts and time to solution when using generated hints?
% \end{enumerate}

The contributions of this work are as follows:
% \begin{itemize}
    (1) We design programming assignment hints with different level of cognitive load, motivated by previous work of preferences for different proficiency levels. We show with experiments in an online programming course that LLM-generated hints lead to faster time to solution, with simple hints being more efficient.
    (2) We evaluate the performance of students of different programming proficiency when presented with the different hint types.
    (3) We show that new content, released after the LLM knowledge cut-off date, affects the quality of the hints negatively.
% \end{itemize}
\section{Background}

\subsection{Programming Feedback}
Automated assessment systems usually provide feedback as test-case results, compiler diagnostics, or static-analysis warnings. This feedback is often difficult for novice programmers to interpret, especially when error messages rely on professional terminology or assume prior knowledge \cite{becker2016effective,messer2024grading}. Enhanced compiler error messages can improve readability, but prior work reports mixed effects on learning outcomes and notes challenges in scaling such approaches across courses and languages \cite{10.1145/3231711}.
Human instructors can provide contextualized feedback adapted to a student’s understanding, but this does not scale to large courses. 
Large language models (LLMs) have therefore been explored for automatic feedback generation. Classroom studies report that LLM-generated hints are often perceived as more readable than compiler output and can support student progress across repeated submissions, while also revealing potential risks of over-reliance on generated feedback \cite{RN5981,10.1145/3478431.3499372}.
Other evaluations report that LLM-generated feedback may include overly detailed guidance, complete solutions, or hallucinated issues, and that students often use such feedback primarily for immediate help-seeking rather than deeper self-regulatory strategies \cite{xiao2024exploring,RN6178}.
Designing LLM-based feedback around pedagogical principles influences how students engage with generated hints, with planning-oriented support showing stronger associations with performance than other forms of assistance \cite{RN6189}.
Instructor-based evaluations also indicate that the quality and usefulness of LLM-generated programming feedback depend strongly on prompting strategies and task difficulty, with substantial variation observed across zero-shot, one-shot, and few-shot configurations \cite{RN6198}.

\subsection{Novice vs. Advanced Programmers}
Programming behavior differs across experience levels. Novice programmers spend more time debugging, make more syntax-related errors, and rely more on trial-and-error approaches. More experienced programmers make fewer errors and perform more targeted code modifications \cite{chuang2024analyzing,shrestha2022pausing}. These differences are reported when proficiency is defined by prior experience rather than task performance \cite{chuang2024analyzing}.
The \emph{expertise reversal effect} describes how instructional approaches that benefit learners with low prior knowledge may become less effective as expertise increases \cite{kalyuga2007expertise,kalyuga2009expertise}. This effect has been discussed in programming education in relation to the usefulness of detailed guidance at different experience levels.
More broadly, adapting the amount and form of instructional explanation to user expertise has been shown to improve learning outcomes in adaptive systems, suggesting that differences in feedback presentation across proficiency levels may be beneficial \cite{RN553}.
Studies of automated program repair report that novice programmers often struggle to interpret repair suggestions, which may differ substantially from their original solutions, while more experienced learners report greater usefulness; students also tend to value error localization over full repair suggestions \cite{kurniawan2023helpful}.
These studies also report that students often value information about the location of an error and that full repair suggestions can be difficult to apply \cite{kurniawan2023helpful}.
Our work examines whether LLM-generated feedback that differs by programming proficiency and explanation depth is associated with differences in student performance.

\section{Methodology}

\subsection{Context}

We applied LLM-generated feedback on an online introductory Web Development course in the span of two months. When a student submits incorrect assignment solutions to the system, by default they see compiler error messages. We assigned at random for part of the users, one of the types of feedback messages. There was an indication that the feedback was AI-generated, and the students had the option to vote for each message as \votegood{} or \votebad.

\subsection{Prompt Design}

We used three types of prompts (\autoref{sec:appendix:prompts}), adjusted for different proficiency levels.
% \begin{itemize}
    \hintgeneral{}: This prompt generates up to three issues in the code, each of which contains description, explanation and hints for next steps.
    \hintnovice{}: This prompt contains only one error, with its description, explanation and guidance for next steps towards the resolution. This is motivated by prior work which states that novice programmers could be overwhelmed by the feedback complexity \cite{xiao2024exploring}. 
    \hintadvanced{}: This prompt contains up to three issues with the code, but gives only descriptions, not hints for how to address the errors. This structure is motivated by the expertise reversal effect \cite{kalyuga2009expertise}, where more advanced programmers perform better with more limited guidance.
% \end{itemize}
We used GPT-4o for generating all feedback messages.

% \input{32-prompts}

% Examples for each of the three types of feedback are shown in \autoref{sec:appendix:feedback_examples}.
% \autoref{fig:feedback-examples}.

% \input{33-feedback-examples}

% We refined the prompts used in the course through several steps.
% Initially, we used one prompt for all types of hints, and only changed the mentioned level. The resulting hints did not differ much.
% Then, we included in the prompt explanations of preferences of novice and advanced programmers from literature. These prompts resulted in more diverse hints for the different levels, but still did not differ significantly.
% Then, we explicitly mentioned the type of content that the hints should include, and ended up having a different prompt for the different hint types. We use GPT-4o for generating all hints.

\subsection{Data}

The data used in this study is from an online introductory Web Development course. We only used submissions of students who allowed using their data for research. 
We excluded all exercise-user pairs for which at least one submission is longer than 2000 characters -- to keep the prompt length shorter, no feedback is displayed for such submissions.
We had three types of feedback, and one control group, (\hintnohint), which did not see any LLM-generated messages. In addition to the LLM-generated feedback, the students also saw the errors logs from the failed automated tests.

We used the data from two parts of the course -- \textit{Part 2: Client-Side Development} and \textit{Part 3: Server-Side Development}. Part 2 contains materials about more recent technologies, which were released after the used model cutoff date, and we refer to it as \coursepartnewcontent. Therefore we also examined the results per course type, to explore whether the recency of the materials affected the generated hints. We refer to Part 3, which contains content that was used in the model training data, and we refer to it as \coursepartoldcontent.

We focused our investigation only on submissions for which the students were actively focused and worked on the task (i.e., engaged). Previous work defines that if a student does not do any input on the programming task for more than 10 minutes, they are likely to be disengaged \cite{shrestha2022pausing}. 
Combining this insight together with the distribution of the number of exercises that were solved in particular time intervals, we analyzed assignments that were completed by a student for up to 30 minutes, as nearly 95\% of all assignments were solved is in this time interval.

Before starting the course, the students were asked the following question about their proficiency level: \textit{Prior programming experience (1=less than 1 year, 2 = 1-2 years, 3 = 2-3 years, 4=3-5 years, 5=6-10 years, 6=more than 10 years)}, which they were not required to answer. 
% We also had questions that asked about the student's self-assessed proficiency on a scale form 1 to 7, but we noticed that the answers to this question do not correlate to the experience in years -- the students were much more optimistic about their skill level. 
Following previous research \cite{chuang2024analyzing}, we split the students as \novicestudents{} if their experience was below 3 years (answers 1, 2 or 3 from the above question), and as \advancedstudents{} if they had over three years of experience (answers 4, 5 or 6 to the above question). When using this separation, there were 111 \novicestudents{} and 38 \advancedstudents{}.

After filtering out the submissions as described above, we obtained a dataset with 3153 assignment-user pairs, of which 1228 are \hintnohint, 665 are \hintgeneral, 639 are \hintnovice, 621 are \hintadvanced.

% The descriptive statistics per feedback type are shown in \autoref{tab:summary_outcomes_and_hint_metrics}.
% In our current setup, each user sees one unique hint type for each assignment that they attempt, but they can see different hint type for the different assignments.
% This table also shows descriptive statistics about the hint length and complexity, and the weighted Levenshtein distance between the change in two consecutive submissions. 
% The three feedback types differ in their textual properties: \hintgeneral{} are substantially longer and more variable in length, \hintnovice{} are shorter and the easiest to read, and \hintadvanced{} are similarly short but more linguistically complex. The average code change between consecutive submissions is similar across hint-based conditions, with \hintnovice{} resulting in the smallest code changes, which aligns with their focus on addressing one issue at a time. The \hintnohint{} baseline shows slightly larger and more variable code changes.
% Very similar solution rate was observed for all the feedback types, as well as the case where only compiler errors were shown, therefore the different feedback types did not affect whether the students solve the assignment.
% The solution rates per hint type are shown in \autoref{tab:solve_rates_by_hint}, and show that the hint does not affect whether the students solve the assignment.

\subsection{Approach}

\subsubsection{Modeling approach}

To answer the three research questions, we conducted three experiments, each corresponding to one research question. 
All analyses were performed on \emph{assignment--user pairs} that were eventually solved, considering submission sequences of up to 30 minutes, as motivated above.
Each student sees the same feedback type for all their submissions to the same assignment.
For each assignment attempted by the student, we assign the feedback type at random. This design choice was made, in order to not put any student in disadvantage of seeing only one feedback type, before we had any information about the quality of each type.

Because each student may submit multiple solutions to the same assignment and may receive different hints for the different assignments, observations cannot be assumed to be independent.
We therefore employ mixed-effects models \cite{lindstrom1988newton,brysbaert2018power}, which allow us to account for repeated measures at both the student and assignment levels.
Across all experiments, we include fixed effects for hint type and, where relevant, student level or course part, as well as random intercepts for students and assignments.
In addition, we include random slopes for hint type at the student level, allowing the effect of hints to vary across students.

For time-based outcomes, we use linear mixed-effects models (LMMs) \cite{lindstrom1988newton, brysbaert2018power} on log-transformed time.
For count-based outcomes (number of attempts), we use negative binomial generalized linear mixed-effects models (NB-GLMMs) \cite{breslow1993approximate, hilbe2011negative} with a log link to account for overdispersion.
Linear mixed-effects models were fit using \texttt{statsmodels} version 0.14 \cite{seabold2010statsmodels}.
Negative binomial generalized linear mixed-effects models were fit using \texttt{Bambi} version 0.17 \cite{capretto2022bambi}.

\subsubsection{Outcome variables}

Each assignment--user pair contains the complete submission sequence up to the first correct solution.
From this sequence, we derive two performance measures:
% \begin{itemize}
    \textbf{\timetosuccess{}}: the time elapsed (in seconds) between the first submission and the first correct submission. 
    Due to right skew, this variable is log-transformed for modeling.
    \textbf{\attemptstosuccess{}}: the number of submissions required until the first correct solution.
    This variable is discrete and overdispersed and is therefore modeled using a negative binomial distribution.
% \end{itemize}

\subsubsection{General model structure}

Let $y_{ij}$ denote the outcome for student $i$ on assignment $j$, and let $\eta_{ij}$ denote the linear predictor:
\begin{align}
\eta_{ij}
&=
\beta_0
+ \boldsymbol{\beta}^{\top}\mathbf{X}_{ij}
+ u_{0i}
+ u_{1i}\,\textit{HintType}_{ij}
+ v_{0j},
\end{align}
where $\mathbf{X}_{ij}$ contains the fixed effects relevant to the experiment,
$u_{0i}$ is a student-specific random intercept,
$u_{1i}$ is a student-specific random slope for hint type,
and $v_{0j}$ is an assignment-specific random intercept.

For \timetosuccess{}, we model:
\begin{align}
\log(y_{ij}) &= \eta_{ij} + \varepsilon_{ij},
&
\varepsilon_{ij} &\sim \mathcal{N}(0,\sigma^2).
\end{align}

For \attemptstosuccess{}, we model:
\begin{align}
y_{ij} &\sim \text{NegBin}(\mu_{ij},\alpha),
&
\log(\mu_{ij}) &= \eta_{ij},
\end{align}
where $\alpha > 0$ is the overdispersion parameter.
The experiments below differ only in the fixed effects included in $\mathbf{X}_{ij}$.

\subsubsection{Experiment 1: Overall effect of feedback on performance (RQ1)}

To address RQ1, we include only hint type as a fixed effect:
\begin{align}
\log(y_{ij})
&=
\beta_0
+ \beta_1\,\textit{HintType}_{ij}
\nonumber\\
&\quad
+ u_{0i}
+ u_{1i}\,\textit{HintType}_{ij}
+ v_{0j}
+ \varepsilon_{ij}.
\end{align}
Here, $\beta_0$ represents the baseline performance under the no-hint condition, and $\beta_1$ captures the average effect of hint type.
The same linear predictor is used for the NB-GLMM when modeling \attemptstosuccess{}.

\subsubsection{Experiment 2: Effect of feedback by student expertise (RQ2)}

To address RQ2, students are categorized as \textit{novice} or \textit{advanced}, and we include an interaction between hint type and student level:

\begin{align}
\log(y_{ij})
&=
\beta_0
+ \beta_1\,\textit{HintType}_{ij}
\nonumber\\
&\quad
+ \beta_2\,\textit{StudentLevel}_i
\nonumber\\
&\quad
+ \beta_3\,(\textit{HintType}_{ij}\times\textit{StudentLevel}_i)
\nonumber\\
&\quad
+ u_{0i}
+ u_{1i}\,\textit{HintType}_{ij}
+ v_{0j}
+ \varepsilon_{ij}.
\end{align}
The interaction term $\beta_3$ captures whether the effect of hint type differs between novice and advanced students.
An analogous NB-GLMM with the same linear predictor is used for \attemptstosuccess{}.

\subsubsection{Experiment 3: Effect of feedback across course parts (RQ3)}

To address RQ3, we include course part (used as a proxy for content recency) and its interaction with hint type:

\begin{align}
\log(y_{ij})
&=
\beta_0
+ \beta_1\,\textit{HintType}_{ij}
\nonumber\\
&\quad
+ \beta_2\,\textit{CoursePart}_{ij}
\nonumber\\
&\quad
+ \beta_3\,(\textit{HintType}_{ij}\times\textit{CoursePart}_{ij})
\nonumber\\
&\quad
+ u_{0i}
+ u_{1i}\,\textit{HintType}_{ij}
+ v_{0j}
+ \varepsilon_{ij}.
\end{align}

The interaction term tests whether the effect of hint type differs between earlier and later course parts.
As before, \attemptstosuccess{} is modeled using an NB-GLMM with the same linear predictor and a negative binomial likelihood.

\subsection{Ethics Statement}

The work followed the ethical procedures at Aalto University.
% (Ethics approval number \textit{Anonymized}).
\section{Results}

\subsection{Experiment 1: Effect of hint types on the student performance on the assignment.}
\begin{figure}[ht]
    \centering
    \includegraphics[width=\linewidth, alt={Horizontal coefficient plot showing the fixed effects of hint types on time to success relative to a no-hint baseline. General hints have a small negative effect of approximately −0.12, with confidence intervals extending close to zero. Single-issue hints show a larger negative effect of about −0.21, and diagnostic hints show a similar effect of approximately −0.20, both with confidence intervals clearly below zero. Negative values indicate reduced time to success. A dashed vertical line at zero denotes no effect.}]{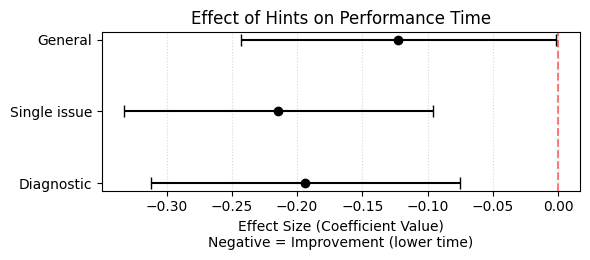}
    \caption{Effects of hint type on \timetosuccess{}, compared to the \hintnohint{} baseline.}
    \label{fig:fixed_effects}
\end{figure}

\paragraph{\timetosuccess{}}
\autoref{fig:fixed_effects} 
summarizes the fixed effects of hint type on log-transformed time to success with the \hintnohint{} condition as the reference level\footnote{Note: In all tables with results, we mark in bold the results that were statistically significant, i.e. the Confidence Interval (CI) for MixedLM or Highest Density Interval (HDI) for NB-GLMM excludes zero. Detailed numbers for time to success are given in the appendix.}.
All hint types were associated with a reduction in completion time relative to the \hintnohint{} baseline. The largest average reduction was observed for \hintnovice{} ($\beta=-0.214$), followed by \hintadvanced{} ($\beta=-0.194$) and \hintgeneral{} ($\beta=-0.122$). These effects are also reflected in the corresponding fixed-effect estimates shown in \autoref{fig:fixed_effects}.
Beyond these average effects, the mixed-effects model accounted for substantial between-student and between-exercise variability. Random intercepts were included for both students and exercises, capturing baseline differences in performance. For time to success, the estimated variance of the student-level random intercept was $\sigma^2_{\text{student}} = 0.103$, while the exercise-level random intercept variance was larger ($\sigma^2_{\text{exercise}} = 1.025$), reflecting notable differences in task difficulty. 
In addition to random intercepts, the models allowed the effect of hint type to vary across students via student-specific random slopes. 
% \autoref{fig:rq1_random_slopes} shows the distribution of these random slope estimates for \timetosuccess{}. 
A wide spread of slopes was observed, which indicates considerable heterogeneity in how individual students responded to hints: while the average effect of hints was beneficial, some students benefited substantially more than others, and a smaller subset showed little or no improvement. This variability suggests that the reported fixed effects capture overall trends, but mask meaningful differences in individual responsiveness to hint support.

% Fixed effects visual

% Fixed effects for time

% \begin{table}
% \caption{Fixed effects for log time to success (RQ1). Reference level: no hint.}
% \label{tab:rq1_fixed_effects_time}
% \scriptsize
% \begin{tabular}{lccl}
% \toprule
%  & \textbf{$\beta$} & \textbf{95\% CI} & \textbf{$p$} \\
% \midrule
% \hintgeneral   & \textbf{-0.122} & \textbf{[-0.243, -0.002]} & \textbf{0.046 *} \\
% \hintnovice    & \textbf{-0.214} & \textbf{[-0.333, -0.096]} & \textbf{0.000 ***} \\
% \hintadvanced  & \textbf{-0.194} & \textbf{[-0.312, -0.075]} & \textbf{0.001 **} \\
% \bottomrule
% \end{tabular}
% \end{table}

% %  Student random slopes
% \begin{figure}[ht!]
%     \centering
%     \includegraphics[width=\linewidth, alt={Scatter plot showing student-level random slope estimates for time to success, ranked from lowest to highest. Each point represents one student's deviation from the average hint effect. Slopes range approximately from −0.45 to +0.37, with a dashed horizontal line at zero indicating no individual deviation. The distribution shows substantial heterogeneity, with some students benefiting strongly from hints (negative slopes) and others showing increased time to success (positive slopes).}]{img/rq1_student_random_slopes_time_to_success.png}
%     \caption{Student random slopes for \timetosuccess{}.}
%     \label{fig:rq1_random_slopes}
% \end{figure}

\paragraph{\attemptstosuccess{}}

The results for attempts, modeled using a NB-GLMM are shown in \autoref{tab:rq1_attempts_nbglmm}. The results indicate that all hint types credibly reduced the number of attempts to success relative to the \hintnohint{} baseline, with the largest reduction observed for \hintadvanced{}. The smallest for improvement is for \hintnovice{}, which could be explained with the fact that this hint type focuses on one issue only, and therefore might lead to a solution after more iterations.
The standard deviation of the exercise-level random intercepts ($\approx 0.26$) reflects variability in baseline attempts across exercises, indicating differences in inherent task difficulty, while the standard deviation of the user-level random intercepts ($\approx 0.17$) indicates more moderate variability in baseline attempts across users. The standard deviations of the user-level random slopes for hint type were comparatively small ($\approx 0.05 - 0.07$), suggesting limited variation across users in how the effect of hints on attempts differs.

\begin{table}
\centering
\caption{Fixed effects of hint type on attempts to success (RQ1).}
\small
\label{tab:rq1_attempts_nbglmm}
\begin{tabular}{lcc}
\toprule
Predictor & $\beta$ & 94\% HDI \\
\midrule
Intercept & 1.216 & [1.130, 1.306] \\
\hintgeneral & \textbf{-0.175} & \textbf{[-0.238, -0.113]} \\
\hintnovice & \textbf{-0.090} & \textbf{[-0.157, -0.030]} \\
\hintadvanced & \textbf{-0.200}& \textbf{[-0.262, -0.134]} \\
\bottomrule
\end{tabular}
\end{table}

\subsection{Experiment 2: Effect of the feedback for students with different expertise.}
\paragraph{\timetosuccess{}}

\begin{figure}[ht]
    \centering
    \includegraphics[width=\linewidth, alt={Line plot with error bars showing predicted time to success across hint conditions (baseline, general, single-issue, diagnostic), separated by student level (novice vs. advanced). Advanced students show a pronounced reduction in time for single-issue hints, while novices show more modest differences across hint types. Error bars indicate uncertainty around the predictions.}]{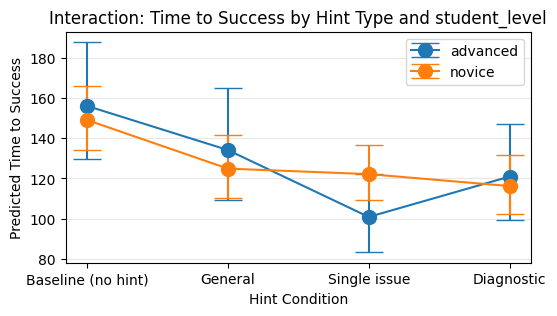}
    \caption{Interaction plot of hint type and student level, with predicted time to success from MixedLM model for assignments solved by up to 30 minutes.}
    \label{fig:interaction_hint_student_level}
\end{figure}

To examine whether the effects of hint type differed between \novicestudents{} and \advancedstudents{}, we fit mixed-effects models including the interaction between hint type and student level for both \timetosuccess{}. \autoref{fig:interaction_hint_student_level} shows the model-predicted outcomes.
% , while the corresponding fixed-effect estimates are reported in \autoref{tab:rq2_fixed_effects_time_interaction}.
\hintnovice{} and \hintadvanced{} are associated with improvement in the time for completion. In contrast, \hintgeneral{} does not show a clear effect. The main effect of student level is not statistically significant once hint type is accounted for.
Although the interaction terms between hint type and student level are not statistically significant, the predicted trends suggest that the relative effectiveness of hint types might differ between \novicestudents{} and \advancedstudents.

%  Interaction plot for hint type x student level

% % Fixed effects on the interaction model - time

% \begin{table}[ht]
% \centering
% \scriptsize
% \caption{Fixed effects for log time to success with hint type × student level interaction (RQ2).}
% \label{tab:rq2_fixed_effects_time_interaction}
% \begin{tabular}{lcc}
% \toprule
%  & \textbf{$\beta$} & \textbf{95\% CI} \\%& \textbf{$p$} \\
% \midrule
% \hintgeneral & -0.151 & [-0.411, 0.108] \\%& 0.252 \\
% \hintnovice & \textbf{-0.436} & \textbf{[-0.697, -0.176]} \\%& \textbf{0.001} ** \\
% \hintadvanced & \textbf{-0.255} & \textbf{[-0.507, -0.002]} \\%& \textbf{0.048} * \\
% \novicestudents & -0.046 & [-0.259, 0.167] \\%& 0.674 \\
% \hintgeneral $\times$ \novicestudents & -0.026 & [-0.327, 0.276] \\%& 0.868 \\
% \hintnovice $\times$ \novicestudents & 0.237 & [-0.064, 0.538] \\%& 0.123 \\
% \hintadvanced $\times$ \novicestudents & 0.006 & [-0.290, 0.302] \\% & 0.969 \\
% \bottomrule
% \end{tabular}
% \end{table}

\paragraph{\attemptstosuccess{}}

\autoref{tab:rq2_attempts_nbglmm} shows the results of the negative binomial mixed-effects model examining the effects of hint type, student level, and their interaction on attempts to success, with \hintnohint{} and \advancedstudents{} as the reference categories.
For \advancedstudents{} (the reference group), all hint types were associated with a reduction in the number of attempts to success relative to \hintnohint{}, with posterior mean effects ranging from $\beta = -0.20$ to $-0.22$ and $94\%$ highest density intervals fully below zero. In contrast, baseline differences between \novicestudents{} and \advancedstudents{} in the \hintnohint{} condition were small and uncertain. 
From these results it seems that \novicestudents{} do not benefit from a particular hint type, and make slightly more submissions before submitting a successful solution when they see \hintnovice.

\begin{table}
\centering
\small
\caption{Fixed effects of hint type, student level, and their interaction on attempts (RQ2).}
\label{tab:rq2_attempts_nbglmm}
\begin{tabular}{p{3.6cm}cc}
\toprule
Predictor & $\beta$ & 94\% HDI \\
\midrule
Intercept & 1.228 & [1.105, 1.334] \\
\hintgeneral{} & \textbf{-0.196} & \textbf{[-0.325, -0.086]} \\
\hintnovice{} & \textbf{-0.220} & \textbf{[-0.339, -0.097]} \\
\hintadvanced & \textbf{-0.207} & \textbf{[-0.319, -0.089]} \\
\novicestudents{} & -0.019 & [-0.117, 0.078] \\
\hintgeneral{} $\times$ \novicestudents{} & 0.027 & [-0.107, 0.168] \\
\hintnovice{} $\times$ \novicestudents{} & \textbf{0.174} & \textbf{[0.030, 0.316]} \\
\hintadvanced $\times$ \novicestudents{} & 0.007 & [-0.132, 0.140] \\
\bottomrule
\end{tabular}
\end{table}

\subsection{Experiment 3: Effect of the content recency on the quality of the LLM-generated feedback.}
\begin{figure}[ht]
    \centering
    \includegraphics[width=\linewidth, alt={Line plot with error bars depicting predicted time to success by hint condition for two course parts (part-2 and part-3). Part-3 students consistently show lower predicted times than part-2 students, with the largest improvement under single-issue hints. Error bars represent uncertainty in the estimates.}]{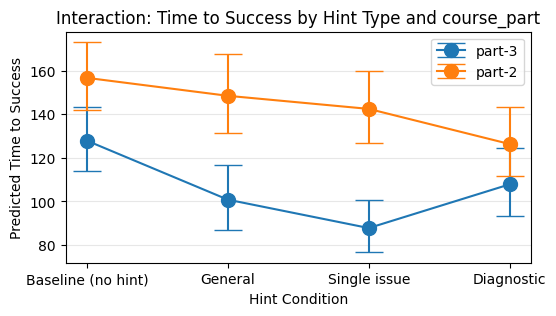}
    \caption{Interaction plot of hint type and course part, with predicted time to success from MixedLM model for assignments solved by up to 30 minutes.}
    \label{fig:interaction_hint_course_part}
\end{figure}

\paragraph{\timetosuccess{}}
To investigate whether the effectiveness of different hint types varies across course progression, we fit mixed-effects model including the interaction between hint type and course part for \timetosuccess{}. Model-predicted outcomes are shown in \autoref{fig:interaction_hint_course_part}.
% , with fixed-effect estimates reported in \autoref{tab:rq3_fixed_effects_time_coursepart_interaction}. 
The reference level corresponds to assignments with \hintnohint{} from \coursepartnewcontent{}.
All hint types for the \coursepartoldcontent{} perform better than \coursepartnewcontent{}. All hint types show improvement over the \hintnohint{} baseline for both course parts, with \hintnovice{} statistically significant for \coursepartoldcontent{}. The \hintadvanced{} performs significantly better for \coursepartnewcontent{}, likely due to the fact that the model could correctly identify errors in the submission, but if it tries to give a direction for fixing, and refers to older version of the material, this could lead to incorrect suggestions.

% % Fixed effects for time - hint type x course part
% \begin{table}
% \centering
% \scriptsize
% \caption{Fixed effects for log time to success with hint type × course part interaction (RQ3).}
% \label{tab:rq3_fixed_effects_time_coursepart_interaction}
% \begin{tabular}{lcc}
% \toprule
%  & \textbf{$\beta$} & \textbf{95\% CI}  \\%& \textbf{$p$} \\
% \midrule
% \hintgeneral & -0.054 & [-0.205, 0.097] \\%& 0.484 \\
% \hintnovice & -0.095 & [-0.245, 0.055] \\%& 0.213 \\
% \hintadvanced & \textbf{-0.215} & \textbf{[-0.368, -0.062]} \\%& 0.006 ** \\
% \coursepartoldcontent & \textbf{-0.203} & \textbf{[-0.339, -0.067]} \\%& 0.003 ** \\
% \hintgeneral $\times$ \coursepartoldcontent & -0.184 & [-0.409, 0.042] \\%& 0.111 \\
% \hintnovice $\times$ \coursepartoldcontent & \textbf{-0.281} & \textbf{[-0.503, -0.059]} \\%& \textbf{0.013} * \\
% \hintadvanced $\times$ \coursepartoldcontent & 0.045 & [-0.181, 0.272] \\%& 0.697 \\
% \bottomrule
% \end{tabular}
% \end{table}

%  Interaction plot for hint type x course part

\paragraph{\attemptstosuccess{}}

The results for attempts, modeled using a negative binomial mixed-effects model and shown in \autoref{tab:rq3_attempts_nbglmm}. Unlike the \timetosuccess, there is no observed significant difference in the \attemptstosuccess{} between the two parts when no LLM-generated feedback is shown. For the \coursepartnewcontent{}, again the \hintadvanced{} is associated with fewer attempts, but interestingly, the \hintgeneral{}, which also hints to solution, shows a reduced number of attempts. For the \coursepartoldcontent{}, the \hintnovice{} is associated with fewer \attemptstosuccess{}.

\begin{table}
\centering
\small
\caption{Fixed effects of hint type, course part, and their interaction on attempts (RQ3).}
\label{tab:rq3_attempts_nbglmm}
\begin{tabular}{p{3.6cm}cc}
\toprule
Predictor & $\beta$ & 94\% HDI \\
\midrule
Intercept & 1.177 & [1.060, 1.292] \\
\hintgeneral{} & \textbf{-0.129} & \textbf{[-0.211, -0.055]} \\
\hintnovice{} & -0.034 & [-0.120, 0.042] \\
\hintadvanced{} & \textbf{-0.217} &\textbf{[-0.300, -0.134]} \\
\coursepartoldcontent{} & 0.089 & [-0.065, 0.257] \\
\hintgeneral{} $\times$ \coursepartoldcontent{} & -0.107 & [-0.225, 0.007] \\
\hintnovice{} $\times$ \coursepartoldcontent{} & \textbf{-0.131} & \textbf{[-0.248, -0.018]} \\
\hintadvanced{} $\times$ \coursepartoldcontent{} & 0.037 & [-0.088, 0.151] \\
\bottomrule
\end{tabular}
\end{table}

\subsection{Qualitative Analysis}

In addition to the quantitative analysis of the student performance in terms of time and attempts to successful submission, we perform qualitative analysis, to identify common errors in the feedback generation. 
When the feedback was displayed to the students, it was accompanied by a message indicating LLM-generated content, and with option to mark the message as \votegood{} or \votebad. Only very few of the feedback messages were evaluated.  Most of the feedback that the users voted for was from Course Part 2 (\coursepartnewcontent{}). There were only two \votegood{} and two \votebad{} voted exercises for Course Part 3 (\coursepartoldcontent). 
The statistics of the given votes are shown in \autoref{tab:vote_stats_full}. The results in the table exclude one \votegood{} exercise that was not solved, and one \votegood{} exercise for which the correct solution was submitted 35 days after the first submission.
Non-surprisingly, when \votegood{} feedback is given, the assignment is solved in a much fewer attempts and for much shorter time than when \votebad{} feedback is shown. Students gave the positive vote in the first appearance of the feedback, and they waited for a few submissions before giving negative vote.
\autoref{tab:hint_type_vote_counts} shows how many votes of each kind were given for the different types of feedback, but because of the small numbers, we cannot make any general conclusions from them.

\begin{table}[h!]
\caption{Number of \votegood{} and \votebad{} votes by feedback type.}
\centering
\small
\begin{tabular}{l r r r}
\toprule
\textbf{Hint Type} 
& \votegood{} 
& \votebad
& \textbf{Total} \\
\midrule
\hintgeneral    & 5  &  8 & 13 \\
\hintnovice     & 7  & 12 & 19 \\
\hintadvanced   & 2  & 11 & 13 \\
\midrule
\textbf{Total}  & 14 & 31 & 45 \\
\bottomrule
\end{tabular}
\label{tab:hint_type_vote_counts}
\end{table}

\begin{table*}[ht]
\centering
\small
\caption{Characteristics of feedback voted as \votegood{} or \votebad{} by students. Attempts and time are measured until the first correct submission. Vote position indicates the submission index at which the vote was given.}
\begin{tabular}{l r r cc cc cc}
\toprule
\textbf{Vote}
& \textbf{\#Pairs}
& \textbf{\#Votes}
& \multicolumn{2}{c}{\textbf{Attempts to Solve}}
& \multicolumn{2}{c}{\textbf{Time to Solve (s)}}
& \multicolumn{2}{c}{\textbf{Vote Position}} \\
\cmidrule(lr){4-5}
\cmidrule(lr){6-7}
\cmidrule(lr){8-9}
& & 
& \textbf{Mean $\pm$ SD} & \textbf{Median}
& \textbf{Mean $\pm$ SD} & \textbf{Median}
& \textbf{Mean $\pm$ SD} & \textbf{Median} \\
\midrule
\votegood
& 13
& 14
& $2.92 \pm 1.04$
& 3
& $279.10 \pm 463.97$
& 155.82
& $1.65 \pm 1.14$
& 1 \\
\votebad
& 22
& 31
& $13.95 \pm 17.34$
& 7.5
& $51288.33 \pm 87726.14$
& 997.92
& $19.45 \pm 20.71$
& 8 \\
\bottomrule
\end{tabular}
\label{tab:vote_stats_full}
\end{table*}

We further perform thematic analysis of the \votebad{} submission, in order to identify common errors. We first went through all the submissions voted as \votebad{} and read them carefully in the context of the assignment description, correct submission and student submission. We also take into account the other submissions by the same user and the same assignment. Then, we coded each voted submission with a description of the type of error it contains. After this, we summarized the codes into common themes. We marked some feedback messages with more than one code, for example in cases where the LLM was referring to an old version, and having had access to the examples in the materials could have showed the expected correct implementation.
Several patterns were observed:
    The suggested fixes were referring to an older version of the taught technology (n=12).
    The pointed issues could be considered meaningful, but examples in the materials were implementing the functionality in the same way as in the student submission, so there was no actual error that the LLM reported (n=12).
    The LLM found no issue with the submission, although the tests were failing (n=7).
    Cases where a wrong issue was proposed, and the real issue in the code was missed (n=4).
    The feedback was actually correct and the student applied it in their correct submission (n=3).

% Most common errors were that the LLM pointed out a problem which was not a real problem, or missed the real issue these. There were several cases where 

\section{Discussion}

Across the course, students receiving LLM-generated feedback reached correct solutions faster than those with compiler output alone, across all feedback types. Differences appear mainly in efficiency (time and attempts), not in final success, as solution rates are high in all conditions.
Clear differences emerge between feedback structures. \hintnovice{} is associated with faster completion but not consistently fewer attempts, suggesting quicker progress without reducing submissions. \hintadvanced{} leads to both fewer attempts and faster completion, indicating more efficient iteration, likely by helping students localize issues. \hintgeneral{} improves performance over no feedback but shows weaker effects than the more constrained formats.
Overall, more detailed guidance does not necessarily improve short-term performance. Limiting feedback to fewer issues or problem identification alone was generally associated with more efficient progress.
When accounting for student level, the models do not show strong interaction effects between feedback type and expertise. However, the predicted outcomes suggest that the relative ranking of feedback types differs between novice and advanced students, with feedback appearing more beneficial for \advancedstudents{} than for \novicestudents{} in our experiments.
These results indicate potential differences in how students at different levels respond to feedback structure, but they do not provide strong evidence for level-specific optimal feedback.
The effectiveness of feedback varied across the different course parts. The results suggest that feedback formats that rely on specific guidance may be more sensitive to limitations in the model coverage of newer technologies.
% Feedback that provides explicit next-step guidance shows stronger effects on course material that was present in the model training data, and weaker effects on newer material that was not.
% This pattern suggests that feedback formats that rely on specific guidance may be more sensitive to limitations in the model’s coverage of newer technologies. 
% Feedback that focuses on identifying issues rather than prescribing fixes appears less affected by differences in course content. This might be related to the fact that the model could correctly identify the issue in the code, but when relying on older version of the technology, could give wrong instructions for solving it.
The analyses show substantial variation across students and assignments. Individual students respond differently to the same feedback type, and assignment difficulty contributes strongly to performance differences. As a result, the reported effects represent average trends rather than uniform improvements.
This indicates that a single feedback strategy is unlikely to be optimal for all students or tasks. The results motivate further work on adaptive feedback approaches that adjust feedback structure based on student behavior, task characteristics, or prior interaction patterns, rather than relying on fixed feedback formats.
Finally, we analyzed LLM-generated feedback evaluated by students and qualitatively examined messages marked as \votebad{}. Due to the small number of votes relative to total submissions, strong conclusions cannot be drawn. Nevertheless, observed failure cases indicate that hallucinations can reduce feedback quality. Providing relevant learning materials in the prompt may mitigate some issues by adding context, especially when the model’s knowledge is outdated. Including compiler error messages could also help guide the LLM toward key issues in the submission.

\section{Conclusion and Future Work}
This study examines how different structures of LLM-generated programming feedback affect student performance in an online course. All feedback types help students reach correct solutions faster than compiler output alone, with differences appearing in efficiency (time and attempts) rather than final success.
Feedback structure influences progress: lower cognitive load (fewer issues or less guidance) leads to more efficient short-term problem solving. However, large individual differences between students outweigh effects of expertise level, suggesting that coarse measures like years of experience may be insufficient.
Feedback effectiveness also varies with course content. Guidance-heavy feedback performs less consistently on newer material, likely due to outdated model knowledge, indicating that relying only on pretrained knowledge is insufficient for content-specific feedback.

Future work should examine long-term learning effects and consistent feedback strategies across courses. Given high variability between students, more fine-grained personalization should be explored, including adaptation based on task, history, and behavior. Student proficiency could be assessed more reliably (e.g., via diagnostics), and preferences for feedback detail studied independently of expertise. Integrating retrieval or up-to-date course materials into feedback generation may improve accuracy, especially for new technologies. 
To mitigate the errors caused by LLM hallucinations, automated fact-checking mechanisms could be deployed to verify the correctness of the LLM-generated feedback before it is displayed to the student.
Finally, future studies should evaluate long-term learning outcomes beyond short-term performance.

\section{Ethical Statement}
Displaying of LLM-generated feedback carries all the risks of using output from generative AI. We have taken several precautions to mitigate these risks. First, the current study was conducted after an ethical approval by our university. Second, we are displaying notifications to the students that the displayed messages are LLM-generated. Third, in order to not put in disadvantage students who only see a particular feedback type, we show different feedback types per student.
For the analysis, we only process data from students who have explicitly agreed their data to be used for research.
In this work, we explore performance on the task, which does not necessarily correspond to deep understanding. We are planning a longitudinal study to ensure that the displayed feedback actually improves the long-term learning outcomes for the students.

\section{Limitations}
This study has several limitations.
First, students were exposed to different hint types across assignments, so assignment–user pairs are not independent. This design avoided disadvantaging students before knowing feedback effectiveness, but future work could assign a single hint type per student or systematically vary it to study personal preferences.
Second, student proficiency is based on self-reported experience, which may not accurately reflect actual ability.
Third, the study is conducted in a single course, limiting generalizability, and excludes submissions over 2000 characters, potentially omitting more complex cases.
Finally, we focus on short-term performance rather than long-term learning, which should be addressed in future longitudinal studies.

% This study has several limitations.
% First, we need to acknowledge the limitations in the setup, where each student sees different hint types for the different assignment they attempt (the hint type is the same for all attempts for the same assignment). Because one user can see different hint types, we cannot assume independence of the analyzed assignment-user pairs. We made this design choice to ensure that no student will be at disadvantage of receiving only a feedback type which is not good, before we knew the performance of each feedback type.
% However, in future work each student could be associated with a specific hint type, or different hint types could be deliberately changed to evaluate personal preferences.
% Second, student proficiency is inferred from self-reported experience. Although we used a separation also proposed in previous work \cite{chuang2024analyzing}, we acknowledge that it may not fully reflect actual programming ability.
% Third, the study is conducted within a single course context, limiting generalizability. In order to reduce the prompt size, we did not analyze submissions of over 2000 characters, which could omit interesting insights for longer and more complex problems.
% Finally, we focus on short-term performance measures and do not assess longer-term learning outcomes. We made the choice to focus on performance, and not long-term learning, as a screening of preparation for longitudinal feedback studies, where we would like to measure the students' performance throughout the course.

%ACKNOWLEDGMENTS are optional
\section{Acknowledgments}

This work was supported by Research Council of Finland grants \#356114 and \#367787.

% \bibliographystyle{abbrv}
\bibliography{latex/ref}

\appendix

\section{Prompts and Examples of Feedback Messages}
\label{sec:appendix:prompts}

\autoref{fig:prompts} shows the prompts used for the three feedback types. For choosing these prompts, we experimented with several different variants, and ended up with this formulation, which outputs the most concise feedback messages.
Examples for the three types of feedback messages are shown on \autoref{fig:feedback-examples}.

\begin{figure*}[ht!]
\centering
\small
\begin{minipage}[t]{0.32\textwidth}
\hintgeneral:
\begin{Verbatim}
SYSTEM
You are a programming feedback tutor. Identify the main problems in the student's code and provide short, direct fixes for each.  
Do not show or quote the sample solution.

RULES
- Refer to the student's identifiers.
- Be concise and factual; no filler or repetition.
- Include up to three important issues.
- For each, give a short Problem / Fix / Why block.
- No code snippets or solution hints.
- Prioritize: 1) Spec/I/O issues, 2) Logic/runtime errors, 3) Edge/complexity issues, 4) Style (only if required).
OUTPUT FORMAT
Issue 1:
Problem: ...
Fix: ...
Why: ...

Issue 2:
Problem: ...
Fix: ...
Why: ...
(Up to 3 issues)
CONTEXT
Assignment: {assignment_description}
Starter Code: {starter_code}
Sample Solution: {sample_solution}
Student Code: {student_code}
TASK
Give compact feedback listing up to three issues, each with Problem / Fix / Why.
\end{Verbatim}
\end{minipage}\hfill
\begin{minipage}[t]{0.32\textwidth}
\hintnovice:
\begin{Verbatim}
SYSTEM
You are a supportive programming tutor. Help a beginner fix one main problem in their code.  
Do not show or quote the sample solution.
RULES
- Focus on one clear issue.
- Use the student's own variable and function names.
- Keep it short, kind, and easy to follow.
- No code or sample-solution hints.
OUTPUT FORMAT
Problem: (what's wrong)
Fix: (one simple action)
Why: (what it fixes or changes)
CONTEXT
Assignment: {assignment_description}
Starter Code: {starter_code}
Sample Solution: {sample_solution}
Student Code: {student_code}
TASK
Give one short piece of beginner-friendly feedback following the format above.
\end{Verbatim}
\end{minipage}\hfill
\begin{minipage}[t]{0.32\textwidth}
\hintadvanced:
\begin{Verbatim}
SYSTEM
You are a peer-level code reviewer. Identify the most relevant issues in the student's code precisely and concisely.  
Do not include or hint at the sample solution.
RULES
- Use the student's identifiers when referencing code.
- Be factual and specific; no filler, no praise, no instructions.
- Describe what is wrong, where, and what impact it has.
- Do not restate the assignment or include any code.
- Prioritize: 1) Spec/I/O issues, 2) Logic/runtime flaws, 3) Edge/complexity issues, 4) Style (only if required).
OUTPUT FORMAT
Issues:
- Problem -> Location -> Impact
- (Up to 3 bullets)
CONTEXT
Assignment: {assignment_description}
Starter Code: {starter_code}
Sample Solution: {sample_solution}
Student Code: {student_code}
TASK
Give terse, peer-style feedback listing only the key issues as bullets.
\end{Verbatim}
\end{minipage}

\caption{The three prompt variants used in the study for the different types of feedback.}
\label{fig:prompts}
\end{figure*}

\begin{figure*}[ht]
\centering
\small
\begin{minipage}[t]{0.32\textwidth}
\hintgeneral:
\begin{Verbatim}
Issue 1:
Problem: The `<h1>` element's text is `"Text transformations"` instead of `"Reactive objects"`.
Fix: Change the heading text to `"Reactive objects"`.
Why: The assignment requires an exact main heading.

Issue 2:
Problem: The imported component name (`Transformations.svelte`) does not match the implemented file name (`ReactiveCounter.svelte`).
Fix: Rename the component file or adjust the import so they match.
Why: Consistent naming is needed to correctly render the component.

Issue 3:
Problem: The paragraph for total numbers incorrectly references `numbers.length` instead of `statsState.numbers.length`.
Fix: Use `statsState.numbers.length` in the paragraph.
Why: It ensures the paragraph displays the correct updated value from the reactive object.
\end{Verbatim}
\end{minipage}\hfill
\begin{minipage}[t]{0.32\textwidth}
\hintnovice:
\begin{Verbatim}
Problem: The `if` condition compares `request.method` to `PEEK` without quotes.

Fix: Replace `PEEK` with `"PEEK"` in the condition.

Why: This ensures that the code correctly compares the request method to the string `"PEEK"`, which is necessary for the condition to work as intended.
\end{Verbatim}
\end{minipage}\hfill
\begin{minipage}[t]{0.32\textwidth}
\hintadvanced:
\begin{Verbatim}
Issues:

- Incorrect internal link text -> `<a href="/contact">Contact</a>` in line 16 -> The text should be "Contact page", affecting compliance with assignment specifications.

- Unnecessary list wrapping links -> `<ul>` element wrapping the `<a>` tags in lines 15-17 -> Deviates from instructions, which require the links to appear directly below the paragraph without being in a list.
\end{Verbatim}
\end{minipage}

\caption{Examples for each of the three variants of feedback used in the study.}
\label{fig:feedback-examples}
\end{figure*}

\section{Data Stats}

The distribution of the number of exercises that were solved in particular time intervals is shown in \autoref{tab:time_bins_all_pairs}.

The solution rates per hint type are shown in \autoref{tab:solve_rates_by_hint}, and show that the hint does not affect whether the students solve the assignment.

The descriptive statistics per feedback type are shown in \autoref{tab:summary_outcomes_and_hint_metrics}.
In our current setup, each user sees one unique hint type for each assignment that they attempt, but they can see different hint type for the different assignments.
This table also shows descriptive statistics about the hint length and complexity, and the weighted Levenshtein distance between the change in two consecutive submissions. 
The three feedback types differ in their textual properties: \hintgeneral{} are substantially longer and more variable in length, \hintnovice{} are shorter and the easiest to read, and \hintadvanced{} are similarly short but more linguistically complex. The average code change between consecutive submissions is similar across hint-based conditions, with \hintnovice{} resulting in the smallest code changes, which aligns with their focus on addressing one issue at a time. The \hintnohint{} baseline shows slightly larger and more variable code changes.
Very similar solution rate was observed for all the feedback types, as well as the case where only compiler errors were shown, therefore the different feedback types did not affect whether the students solve the assignment.

% \begin{figure}[ht!]
%     \centering
%     \includegraphics[width=\linewidth]{img/assignment_completed_per_interval.png}
%     \caption{Number of assignments that students completed in different time intervals.}
%     \label{fig:time_bins_all_pairs}
% \end{figure}

\begin{table*}[ht]
\centering
\small
\caption{Number of assignments that students completed in different time intervals.}
\label{tab:time_bins_all_pairs}
\begin{tabular}{lrrrr}
\toprule
\textbf{Interval} & \hintnohint & \hintgeneral & \hintnovice & \hintadvanced \\
\midrule
$\leq$ 10min            & 1068 & 587 & 585 & 563 \\
10-20min           & 128  & 57  & 41  & 51  \\
20-30min          & 32   & 21  & 13  & 7   \\
30min-2h          & 47   & 20  & 15  & 19  \\
2-24h         & 16   & 9   & 12  & 10  \\
24-72h       & 9    & 3   & 2   & 4   \\
$>$ 72h            & 3    & 2   & 6   & 4   \\
\bottomrule
\end{tabular}
\end{table*}

\begin{table*}[ht]
\centering
\small
\caption{Statistics of the data used in the experiments (assignments that were solved up to 30 minutes). Number of assignment-user pairs, submissions, users, and assignments, together with descriptive properties of the generated hints and subsequent code changes by hint type. Values are reported as mean $\pm$ standard deviation. Hint length is measured in words; hint readability is quantified using Flesch Reading Ease (higher values indicate easier readability); code diff is measured as weighted Levenshtein distance between consecutive submissions. There are 47 assignments in total, and each of the hint types was seen for all assignments.}

\setlength{\tabcolsep}{6pt}
\begin{tabular}{lrrrrrr}
\toprule
\textbf{Hint type} &
{\textbf{Pairs}} &
{\textbf{Submissions}} &
{\textbf{Users}} &
% {\textbf{Assignments}} &
\textbf{Hint Length} &
\textbf{Hint FRE} &
\textbf{Code diff} \\
\midrule

\hintnohint
& 1228
& 4677
& 206
% & 47
& --
& --
& 0.074 $\pm$ 0.186 \\

\hintgeneral
& 665
& 2208
& 161
% & 47
& 88.78 $\pm$ 34.96
& 54.24 $\pm$ 11.12
& 0.071 $\pm$ 0.129 \\

\hintnovice
& 639
& 2312
& 162
% & 47
& 54.23 $\pm$ 10.95
& 57.43 $\pm$ 11.22
& 0.061 $\pm$ 0.129 \\

\hintadvanced
& 621
& 1966
& 162
% & 47
& 56.30 $\pm$ 21.57
& 44.82 $\pm$ 14.42
& 0.072 $\pm$ 0.130 \\

\midrule
Total
& 3153
& 11163
& 223
% & 47
& --
& --
& -- \\
\bottomrule
\end{tabular}
\label{tab:summary_outcomes_and_hint_metrics}
\end{table*}

\begin{table*}[t]
\centering
\small
\caption{Number of assignment--user pairs per feedback type, showing total submissions, solved and unsolved cases, and the percentage of solved submissions.}

\begin{tabular}{lrrrr}
\toprule
\textbf{Feedback type} & \textbf{All} & \textbf{Solved} & \textbf{Unsolved} & \textbf{\% Solved} \\
\midrule
\hintnohint                    & 1328 & 1303 & 25 & 98.1 \\
\hintgeneral              & 708  & 699  & 9  & 98.7 \\
\hintnovice       & 684  & 674  & 10 & 98.5 \\
\hintadvanced     & 663  & 658  & 5  & 99.2 \\
\midrule
\textbf{Total}             & 3383 & 3334 & 49 & 98.6 \\
\bottomrule
\end{tabular}
\label{tab:solve_rates_by_hint}
\end{table*}

\section{Additional Results}

\subsection{Experiment 1 Additional Results:  Effect of hint types on the student performance on the assignment.}

\begin{figure}[ht]
    \centering
    \includegraphics[width=\linewidth, alt={Scatter plot showing student-level random slope estimates for time to success, ranked from lowest to highest. Each point represents one student's deviation from the average hint effect. Slopes range approximately from −0.45 to +0.37, with a dashed horizontal line at zero indicating no individual deviation. The distribution shows substantial heterogeneity, with some students benefiting strongly from hints (negative slopes) and others showing increased time to success (positive slopes).}]{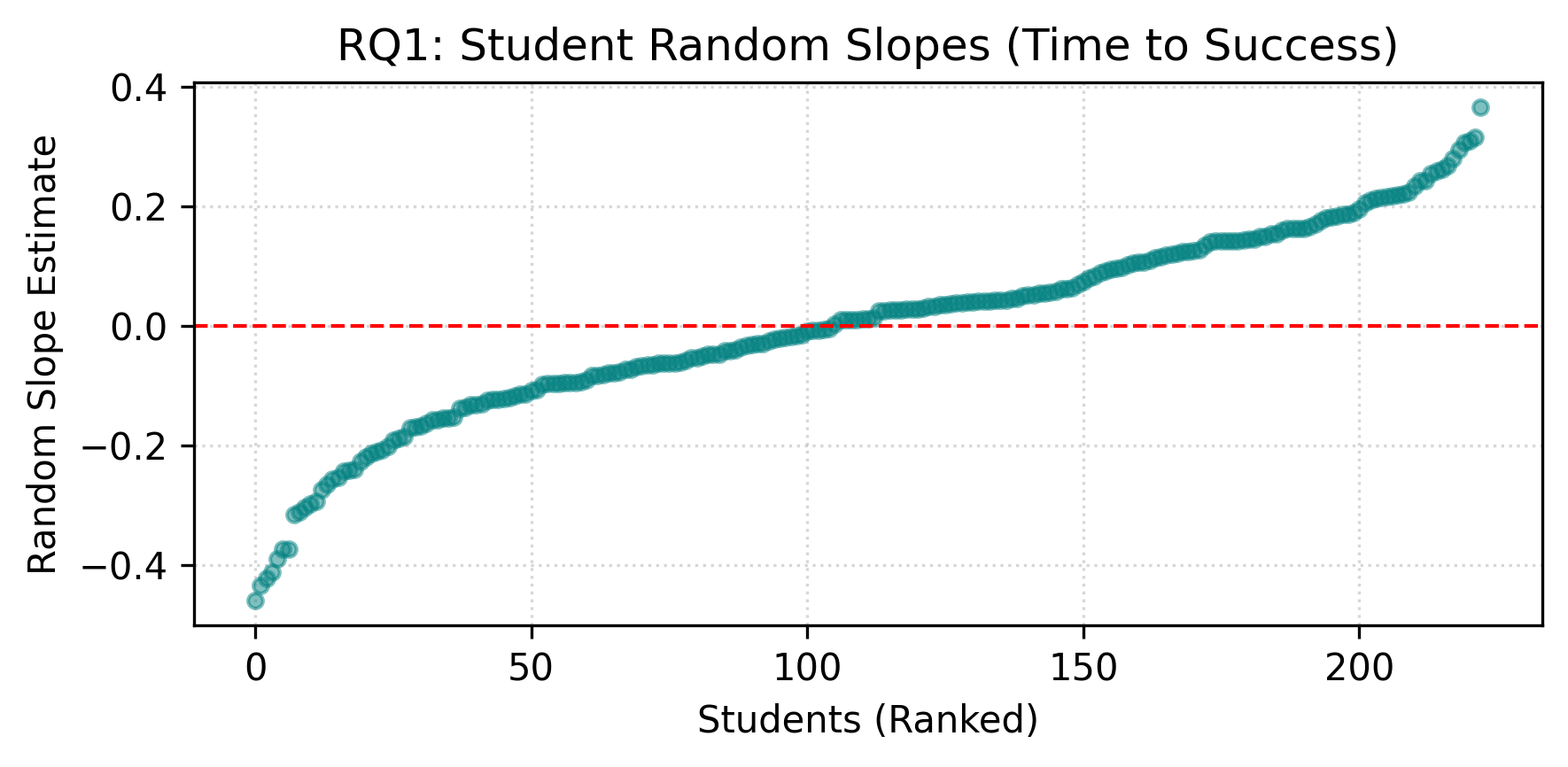}
    \caption{Student random slopes for \timetosuccess{}.}
    \label{fig:rq1_random_slopes}
\end{figure}

\autoref{tab:rq1_fixed_effects_time} shows the fixed effects  of hint type on log-transformed time to success with the \hintnohint{} condition as the reference level.
\autoref{fig:rq1_random_slopes} shows the distribution of the random slope estimates per student for \timetosuccess{}.

% \subsection{Fixed effects for time to success.}

% Fixed effects for time

\begin{table}[ht]
\caption{Fixed effects for log time to success (RQ1). Reference level: \hintnohint.}
\label{tab:rq1_fixed_effects_time}
\small
\begin{tabular}{lccl}
\toprule
 & \textbf{$\beta$} & \textbf{95\% CI} & \textbf{$p$} \\
\midrule
\hintgeneral   & \textbf{-0.122} & \textbf{[-0.243, -0.002]} & \textbf{0.046 *} \\
\hintnovice    & \textbf{-0.214} & \textbf{[-0.333, -0.096]} & \textbf{0.000 ***} \\
\hintadvanced  & \textbf{-0.194} & \textbf{[-0.312, -0.075]} & \textbf{0.001 **} \\
\bottomrule
\end{tabular}
\end{table}

%  Student random slopes

\subsection{Experiment 2 Additional Results: Effect of the feedback for students with different expertise.}

The fixed-effect estimates for log time to success with hint
type × student level interaction are reported in \autoref{tab:rq2_fixed_effects_time_interaction}.

% Fixed effects on the interaction model - time

\begin{table}[ht]
\centering
\small
\caption{Fixed effects for log time to success with hint type × student level interaction (RQ2).}
\label{tab:rq2_fixed_effects_time_interaction}
\begin{tabular}{p{3.6cm}cc}
\toprule
 & \textbf{$\beta$} & \textbf{95\% CI} \\%& \textbf{$p$} \\
\midrule
\hintgeneral & -0.151 & [-0.411, 0.108] \\%& 0.252 \\
\hintnovice & \textbf{-0.436} & \textbf{[-0.697, -0.176]} \\%& \textbf{0.001} ** \\
\hintadvanced & \textbf{-0.255} & \textbf{[-0.507, -0.002]} \\%& \textbf{0.048} * \\
\novicestudents & -0.046 & [-0.259, 0.167] \\%& 0.674 \\
\hintgeneral $\times$ \novicestudents & -0.026 & [-0.327, 0.276] \\%& 0.868 \\
\hintnovice $\times$ \novicestudents & 0.237 & [-0.064, 0.538] \\%& 0.123 \\
\hintadvanced $\times$ \novicestudents & 0.006 & [-0.290, 0.302] \\% & 0.969 \\
\bottomrule
\end{tabular}
\end{table}

\subsection{Experiment 3 Additional Results: Effect of the content recency on the quality of the LLM-generated feedback.}

The fixed-effect estimates for log time to success with hint type × course part interaction are shown in \autoref{tab:rq3_fixed_effects_time_coursepart_interaction}.

% Fixed effects for time - hint type x course part
\begin{table}[ht]
\centering
\small
\caption{Fixed effects for log time to success with hint type × course part interaction (RQ3).}
\label{tab:rq3_fixed_effects_time_coursepart_interaction}
\begin{tabular}{p{3.6cm}cc}
\toprule
 & \textbf{$\beta$} & \textbf{95\% CI}  \\%& \textbf{$p$} \\
\midrule
\hintgeneral & -0.054 & [-0.205, 0.097] \\%& 0.484 \\
\hintnovice & -0.095 & [-0.245, 0.055] \\%& 0.213 \\
\hintadvanced & \textbf{-0.215} & \textbf{[-0.368, -0.062]} \\%& 0.006 ** \\
\coursepartoldcontent & \textbf{-0.203} & \textbf{[-0.339, -0.067]} \\%& 0.003 ** \\
\hintgeneral $\times$ \coursepartoldcontent & -0.184 & [-0.409, 0.042] \\%& 0.111 \\
\hintnovice $\times$ \coursepartoldcontent & \textbf{-0.281} & \textbf{[-0.503, -0.059]} \\%& \textbf{0.013} * \\
\hintadvanced $\times$ \coursepartoldcontent & 0.045 & [-0.181, 0.272] \\%& 0.697 \\
\bottomrule
\end{tabular}
\end{table}

\end{document}